\documentclass[pdftex,twocolumn,epjc3]{svjour3}          

\RequirePackage[T1]{fontenc}

\smartqed  

\RequirePackage{graphicx}
\usepackage{subcaption}
\RequirePackage{mathptmx}      
\RequirePackage{flushend}
\RequirePackage[numbers,sort&compress]{natbib}
\RequirePackage[colorlinks,citecolor=blue,urlcolor=blue,linkcolor=blue]{hyperref}
\usepackage{amsmath}
\usepackage{url}

\journalname{Eur. Phys. J. C}

\begin{document}

\title{Tidal contributions to neutron star orbital precession around Sgr $A^*$}


\author{H Lalrinfela\thanksref{e1,addr1}
        \and
        P.C. Lalremruati\thanksref{e2,addr2}
        \and
        Zodinmawia\thanksref{e3,addr1}
        \and
        N Malsawmtluangi\thanksref{e4,addr2}
}

\thankstext{e1}{e-mail: felablues10@gmail.com}
\thankstext{e2}{e-mail: pcremruati905@gmail.com}
\thankstext{e3}{e-mail: zodinmawia@mzu.edu.in}
\thankstext{e4}{e-mail: tei.naulak@gmail.com}

\institute{Department of Physics, Mizoram University, Tanhril, Aizawl, Mizoram 796004, India\label{addr1}
          \and
          Department of Physics, Govt. Kolasib College, Kolasib, Mizoram 796081, India\label{addr2}
}

\date{Received: date / Accepted: date}

\maketitle

\begin{abstract}
We investigate tidal contributions to the orbital precession of neutron stars around the supermassive black hole Sgr A*, using stellar configurations described by the Tolman–Oppenheimer–Volkoff equations. We examine the dependence of tidal precession on stellar compactness, pericentre distance, and the quadrupolar tidal Love number \(k_2\), and compare its magnitude with general relativistic orbital precession. The calculated tidal angular shift
decreases strongly with increasing pericentre distance and, for the stellar configurations considered, with increasing compactness. Larger Love numbers enhance the tidal contribution. This behaviour reflects the strong dependence of tidal precession on the stellar-radius-to-pericentre-distance ratio, \((R/r_p)^5\). Despite the enhancement during closer encounters, the predicted tidal shifts remain extremely small, reaching approximately \(3.5\times10^{-18}\,\mu\mathrm{as}\) within the displayed parameter range. The ratio of tidal to general relativistic precession spans approximately \(10^{-33}\)–\(10^{-28}\), indicating that relativistic precession overwhelmingly dominates. Our results suggest that direct detection of the neutron-star tidal contribution through orbital precession around Sgr A* would be exceptionally challenging under the assumptions of the adopted model.
\end{abstract}

\section{Introduction}

The Galactic Centre hosting a supermassive black hole of mass $\approx$ 4.3 $\times$ $10^{6}$$M_{\odot}$, with its rich environment of young stars, provides an ideal laboratory for testing gravity  in strong field regime\citep{r1, r2}. Most of the S-stars discovered near the Galactic Centre are main-sequence stars \citep{r3}. Kim and Davis \cite{r4} have also reported the possibility of detection of a population of thousands of neutron stars residing within 1 parsec of the Galactic Centre. Recently, a millisecond pulsar (MSP) PSR J1744-2946, with a period of P = 8.4 milliseconds, was discovered within  $1^\circ$ of the Galactic Centre environment using the South African MeerKAT telescope \citep{r5}. This discovery hints the presence of a larger population of neutron stars near the Galactic Centre environment.

Neutron stars are the most compact stellar objects known, containing masses of order the Sun's within radii of only about 10-12\,km and reaching densities several times that of nuclear saturation \citep{r6}. The bulk properties: mass, radius, and the pressure-density relation of the matter they contain are set by the equation of state of cold, strongly interacting matter, which remains poorly determined at the highest densities probed in stellar interiors \citep{r7}. The maximum mass a neutron star can support is controlled by the stiffness of the EOS at densities in excess of a few times nuclear saturation density \citep{r7, r8}, and the behaviour of the speed of sound at those densities is itself an open question \citep{r9, r10}; neutron stars therefore provide a unique laboratory for matter under conditions inaccessible on Earth \citep{r8}.

The equilibrium structure of a static, spherically symmetric star follows from the Tolman–Oppenheimer–Volkoff (TOV) equations \citep{r11, r12}, which are integrated together with a barotropic EOS to yield the mass–radius relation and the stellar compactness $C = M/R$ \citep{r13, r14}. Radio timing of binary pulsars has provided precise mass measurements, with several neutron stars firmly in the $1.9$--$2.0\,M_\odot$ range and a maximum mass inferred at $2.21^{+0.10}_{-0.13}\,M_\odot$ ($68\%$ credibility) \citep{r6, r15}. Independent radius constraints have come from X-ray observations with the Neutron star Interior Composition Explorer (NICER): for PSR J0030+0451 the reported mass and radius are $M = 1.34^{+0.15}_{-0.16}\,M_\odot$ and $R = 12.71^{+1.14}_{-1.19}$\,km, while for the massive pulsar PSR J0740+6620 they are $M = 2.072^{+0.067}_{-0.066}\,M_\odot$ and $R = 12.39^{+1.30}_{-1.98}$\,km \citep{r16}. Combined with nuclear theory and with the tidal information from gravitational-wave observations, these measurements have substantially narrowed the allowed EOS space \citep{r17, r18, r19}.

A complementary and particularly clean probe of neutron star structure is the star's tidal response to an external gravitational field. The
quadrupolar tidal Love number $k_2$ \citep{r13} quantifies the static deformation of a self-gravitating body in a tidal field and has been extended to relativistic stars by solving the linearized Einstein equations for a static, even-parity $\ell = 2$ perturbation of the TOV background \citep{r13, r20, r21}. In practice, $k_2$ is obtained by integrating the $\ell = 2$ perturbation equation on the TOV background and matching the interior solution to the exterior Schwarzschild spacetime at the stellar surface \cite{r13, r22, r23}. 

In this work, we investigate the tidal Love number and the tidal precession of neutron stars orbiting Sgr\,A$^*$ using the TOV model. Radius of the neutron star and $k_2$ are obtained numerically by integrating the TOV equations, and the quadrupolar Love number from solving the coupled $\ell = 2$ tidal perturbation equation on that background. The paper is arranged in the following order: section 2 discuss the tidal Love number of neutron stars for polytropic TOV model. Section 3 estimates the tidal precession of neutron stars. In section 4, the estimated results are presented with a brief discussion.

\section{Tidal Love Number}
\label{sec:tidal}

The tidal Love number measures how easily the mass distribution of a star is deformed by an external tidal field \citep{r26}.
Consider a neutron star of mass $M$ and radius $R$ in the field of a
companion of mass $M_{\rm bh}$ at a separation $d\gg R$. Let
$\Phi_{\rm ext}$ denote the companion's potential in the neighbourhood of the star, with the star's own potential removed. Expanding $\Phi_{\rm ext}$ about the stellar centre, the constant term can be removed by the choice of the zero of the potential, and the linear term is a uniform acceleration that moves the
star as a whole in free fall, and neither deforms the star. The leading
contribution that does is the tidal potential,
\begin{align}
  \Phi_{\rm tide}(\mathbf{x})
    &= \Phi_{\rm ext}(\mathbf{x}) - \Phi_{\rm ext}(\mathbf{0})
       - x^i\,\partial_i\Phi_{\rm ext}(\mathbf{0})\\
    &= \tfrac{1}{2}\,\mathcal{E}_{ij}\,x^i x^j + O(x^3),
  \label{eq:Edef}
\end{align}    
where, 
\begin{equation}
    \mathcal{E}_{ij} \equiv \partial_i\partial_j\Phi_{\rm ext}\big|_{\mathbf{x}=0}
\end{equation}
\label{eq:Edef}\\
 defines the applied tidal tensor \citep{r27, r28}. Because the companion's potential is source-free at the location of the star, $\nabla^2\Phi_{\rm ext}=0$, so $\mathcal{E}_{ij}$ is symmetric and trace-free, $\mathcal{E}^i{}_i=0$. It therefore has five independent components and is a pure $\ell=2$ (quadrupolar) tensor. Higher multipoles, such as the octupole term in $O(x^3)$, are suppressed relative to the quadrupole by powers of $R/d$ and are neglected here.

\begin{figure}[htbp]
\centering
\includegraphics[width=1.10\linewidth]{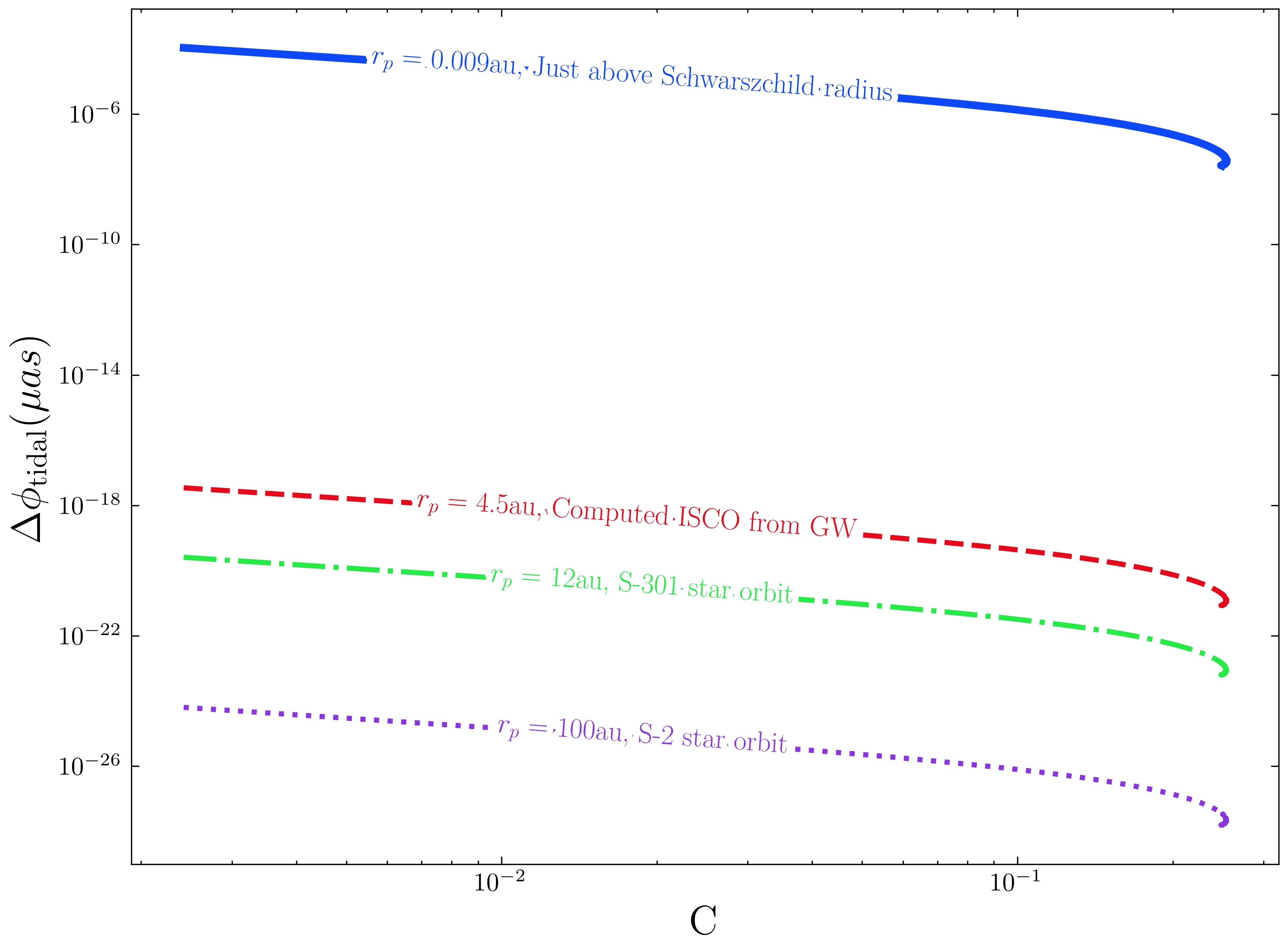}
\caption{Tidal precession as a function of stellar compactness $C$ for
$r_p = 0.009$~au, $4.5$~au, $12$~au, and $100$~au}
\label{fig:tide_compact}
\end{figure}

For a point-mass companion, $\Phi_{\rm ext}=-GM_{\rm bh}/|\mathbf{x}-\mathbf{d}|$,
and Eq.~(\ref{eq:Edef}) gives
\begin{equation}
  \mathcal{E}_{ij}=\frac{GM_{\rm bh}}{d^{3}}\left(\delta_{ij}-3\,n_in_j\right),
  \qquad n_i=\frac{d_i}{d},
  \label{eq:Epoint}
\end{equation}
where $n_i$ is the unit vector from the star to the companion. In a frame whose third axis is aligned with $n_i$, $\mathcal{E}_{ij}=(GM_{\rm bh}/d^{3})\,\mathrm{diag}(-2,1,1)$: the tidal field stretches the star along the line joining the two bodies and compresses it in the transverse directions \citep{r29, r27}. Its magnitude, $|\mathcal{E}|\equiv(\mathcal{E}_{ij}\mathcal{E}^{ij})^{1/2}=\sqrt{6}\,GM_{\rm bh}/d^{3}$, falls off as the inverse cube of the separation. The tidal driving is therefore strongly concentrated near pericentre, and this steep dependence on orbital separation is what makes the tidal precession of Sec.~\ref{sec:TidalPrecession} so sensitive to $r_p$.

\begin{figure*}[h]
\centering
\begin{minipage}[b]{0.49\linewidth}
    \centering
    \includegraphics[width=\linewidth]{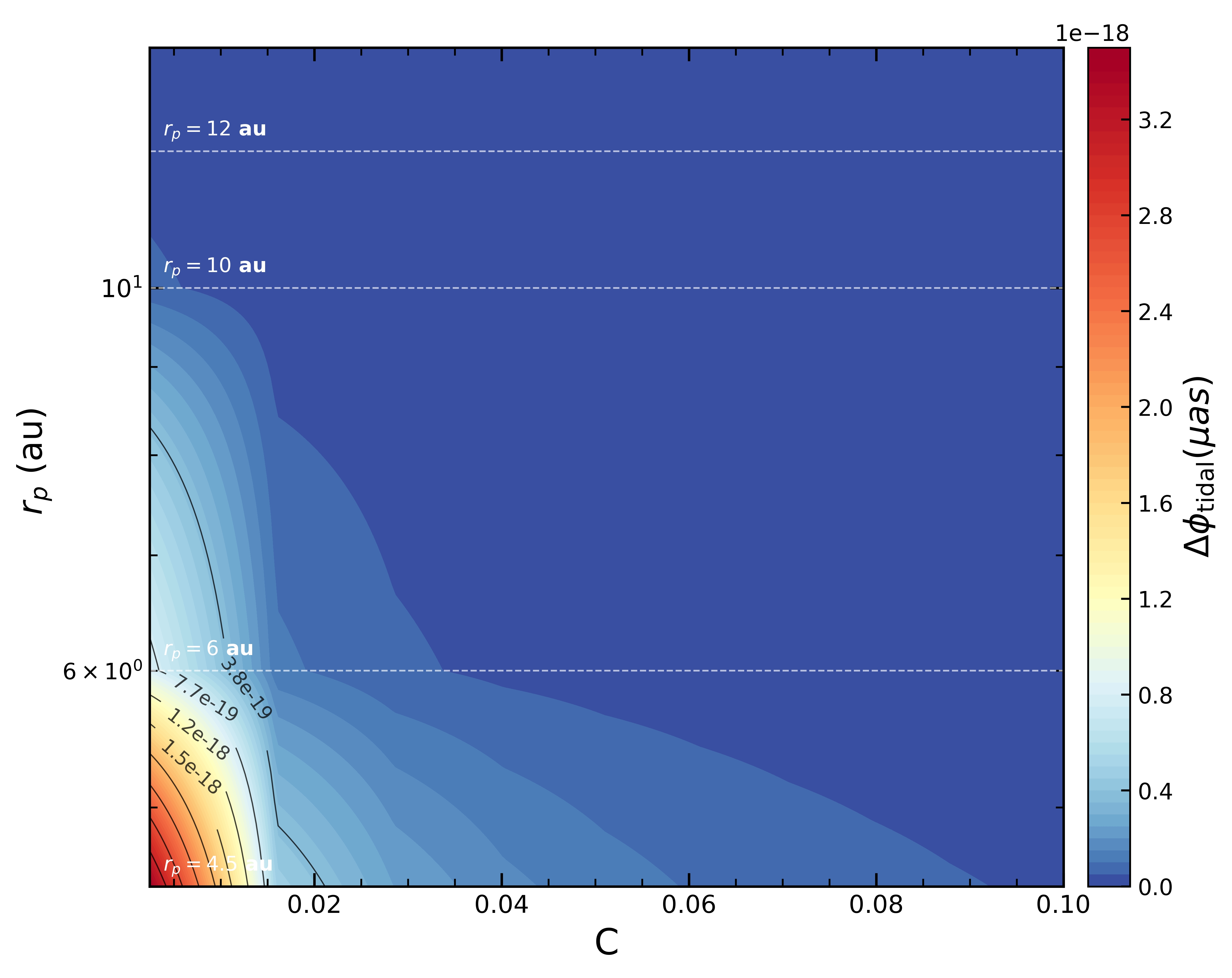}
    \par\small (a)
\end{minipage}
\hfill
\begin{minipage}[b]{0.49\linewidth}
    \centering
    \includegraphics[width=\linewidth]{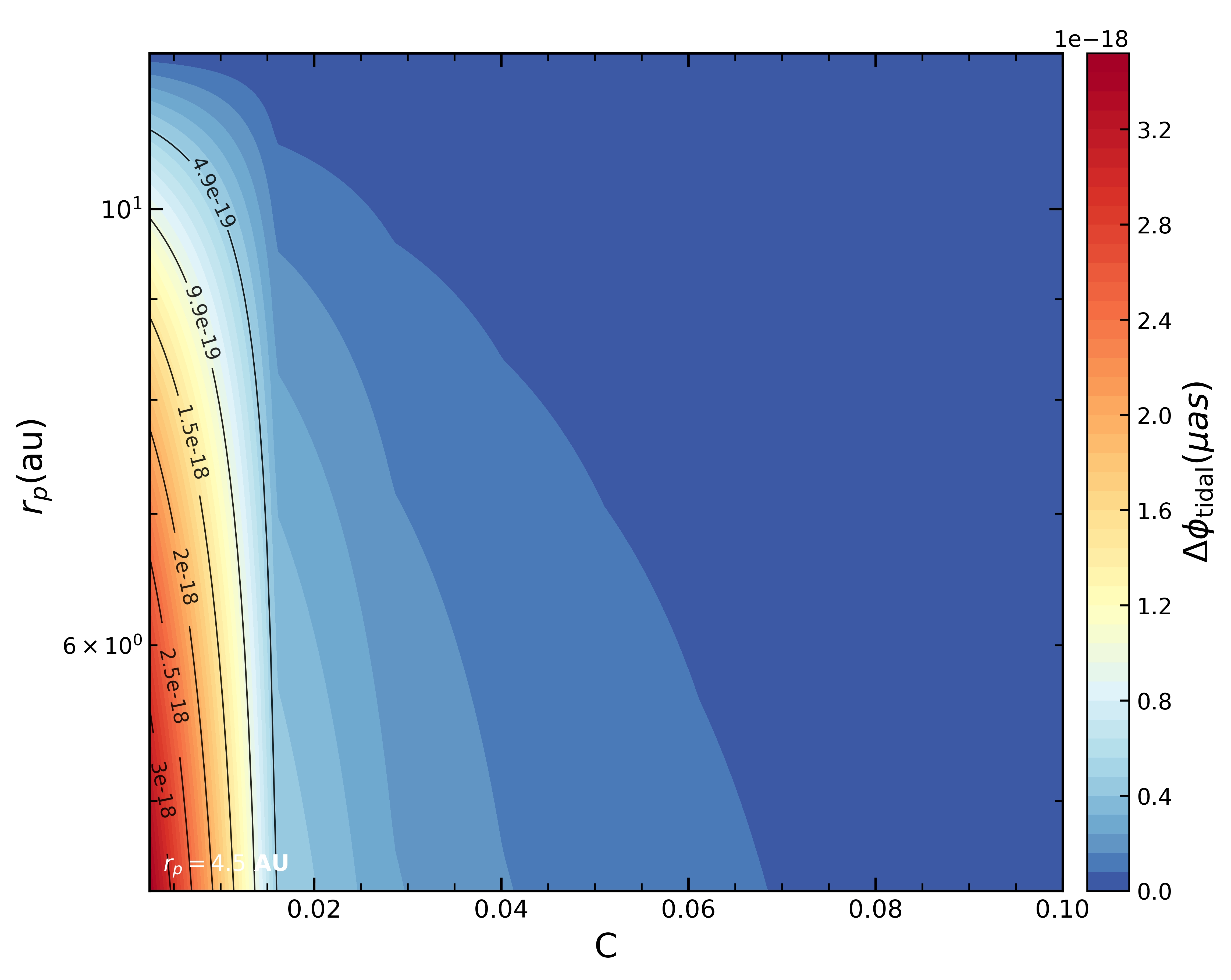}
    \par\small (b)
\end{minipage}
\caption{Pericentre distance $r_p$ versus compactness of a star (C) for various tidal  precession, $\Delta\phi_{\rm tidal}$, of neutron stars measured in micro arcsecs. Figure (a) shows the variation till $r_p$ = 12au and figure (b) highlight the effects of tidal precession within $r_p$ = 4.5au.}
\label{fig:rp_vs_C_a}
\end{figure*}

The response of the star is described by its mass quadrupole moment
\citep{r27},
\begin{equation}
  Q_{ij}=\int \rho(\mathbf{x})\left(x_ix_j-\frac{1}{3}r^{2}\delta_{ij}\right)d^{3}x,
  \label{eq:Qdef}
\end{equation}
which is also symmetric and trace-free. For a small, static deformation, the induced quadrupole is linearly proportional to the applied field
\citep{r28},
\begin{equation}
  Q_{ij}=-\lambda\,\mathcal{E}_{ij},
  \label{eq:response}
\end{equation}
where $\lambda$ is the tidal deformability. The sign is fixed by requiring an elongation towards the companion: with
$\mathcal{E}_{zz}=-2GM_{\rm bh}/d^{3}$ along $n_i$, Eq.~(\ref{eq:response})
gives $Q_{zz}=+2\lambda\,GM_{\rm bh}/d^{3}>0$, as it must. The
linear stimulus response relation between the applied tidal field and the induced quadrupole is the defining property of the Love number formalism, introduced for Newtonian bodies \citep{r30} and extended to full general relativity in Refs.~\citep{r13, r20, r21}. 

The dimensionless quadrupolar Love number $k_2$ is defined through
\begin{equation}
  \lambda=\frac{2}{3}\,\frac{k_2R^{5}}{G},
  \label{eq:lambda}
\end{equation}
where, G is the gravitational constant, R is radius of the star and  $k_2$ is the tidal Love number of a star \cite{r27}.

For this work, we use the tidal Love number obtained from the TOV model given by  \cite{r13} 
\begin{align}
k_2 = \; & \frac{8C^5}{5}(1-2C)^2\left[2 - y + 2C(y-1)\right] \notag \\
& \times \Big\{ 2C\left[6 - 3y + 3C(5y - 8)\right] \notag \\
& \quad + 4C^3\left[13 - 11y + C(3y-2) + 2C^2(1+y)\right] \notag \\
& \quad + 3(1-2C)^2\left[2 - y + 2C(y-1)\right]\ln(1-2C) \Big\}^{-1}.
\label{eq:k2}
\end{align}
where $C=GM/c^2R$ is the compactness of the star, and $y$ is obtained by solving the boundary condition given in \cite{r13}.

\begin{figure*}[t]
\centering
\begin{subfigure}[b]{0.49\textwidth}
    \centering
    \includegraphics[width=\linewidth]{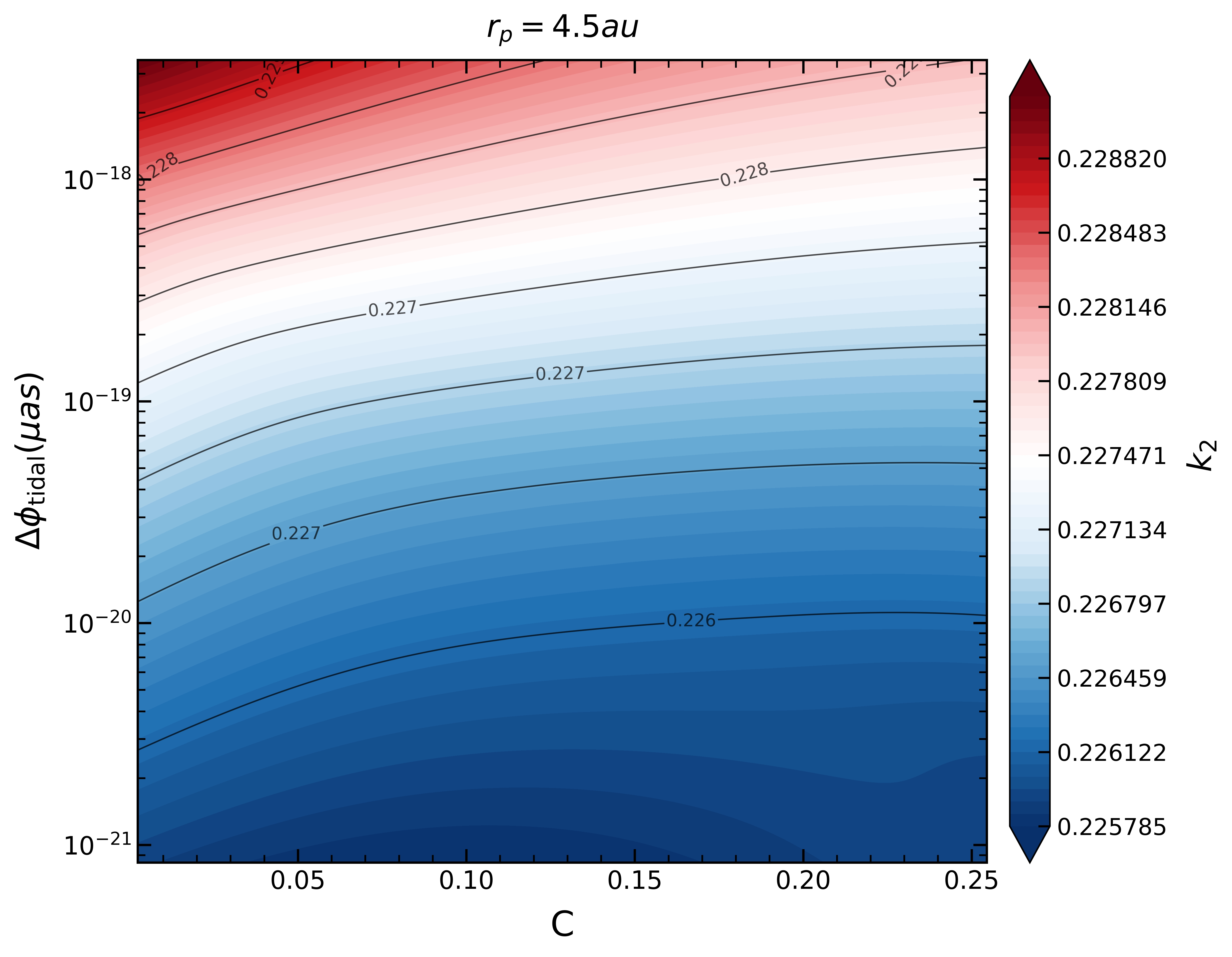}
    \caption{$r_p = 4.5$~au}
    \label{fig:MR_rp4.5}
\end{subfigure}
\hfill
\begin{subfigure}[b]{0.49\textwidth}
    \centering
    \includegraphics[width=\linewidth]{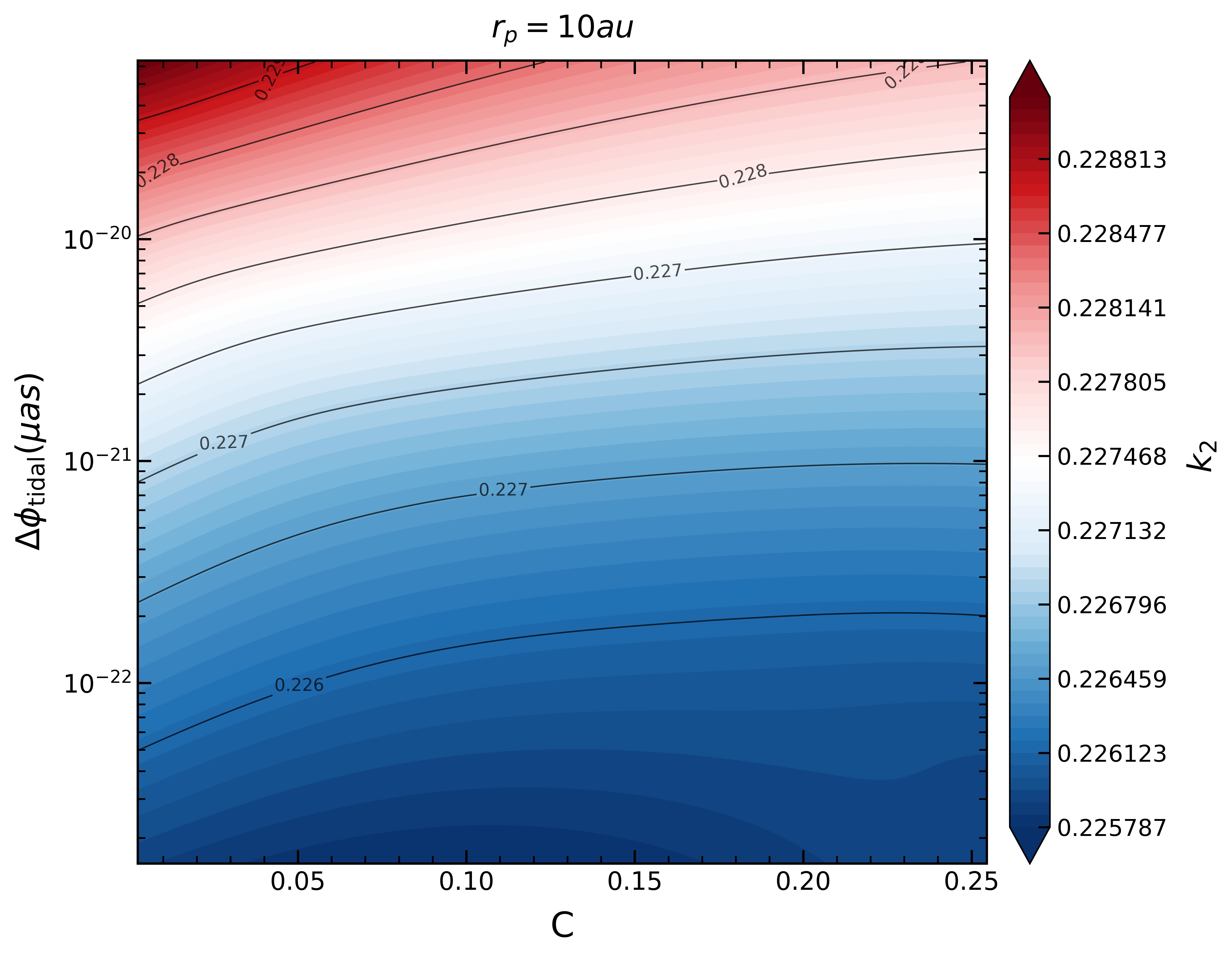}
    \caption{$r_p = 10$~au}
    \label{fig:MR_rp10}
\end{subfigure}
\caption{Variation of tidal precession  for pericentre distances (a)~$r_p = 4.5$~au and (b)~$r_p = 10$~au for different tidal Love number, $k_2$.}
\label{fig:MR}
\end{figure*}

\section{Tidal Precession}
\label{sec:TidalPrecession}

For a neutron star of mass $M$ orbiting a companion of mass $M_{bh}$, the tidal precession per orbit  is given by \cite{r49}

\begin{equation}
\Delta\phi_{\rm tidal} =
\frac{30\pi k_2}{(1+e)^5}
\left(1 + \frac{3}{2}e^2 + \frac{1}{8}e^4\right)
\frac{M_{bh}}{M}
\left(\frac{R}{r_p}\right)^{5},
\label{eq:tide}
\end{equation}
where $M$ and $R$ denote the mass and radius of the neutron star, $M_{bh}$ is the mass of Sgr$A^*$,  $r_p$ is the pericentre distance, and $e$ is the orbital eccentricity.  The companion mass enters through the tidal field strength. Equation~\eqref{eq:tide} is the only expression used to evaluate the tidal precession.

The tidal precession increases with the Love number, 
\begin{equation}
    \Delta\phi_{\rm tidal} \propto k_2,
\end{equation}
and with the mass ratio,

\begin{equation}
    \Delta\phi_{\rm tidal}\propto \frac{M_{\rm bh}}{M}
\end{equation}
Most importantly, it depends on the fifth power of the ratio between the neutron-star radius and orbital separation,

\begin{equation}
    {\Delta\phi_{\rm tidal}\propto \left(\frac{R}{r_p}\right)^5 }
\end{equation}
This fifth-power dependence is responsible for the strong suppression of tidal precession when the orbital separation is large compared with the neutron-star radius. This can be seen in Figure~\ref{fig:tide_compact} and Figure~\ref{fig:MR}.

In this work, we used a highly eccentric orbit, e = 0.9. For highly eccentric orbits, the factor $((1+e)^{-5})$ can become large. Consequently, tidal effects can increase substantially as the orbit becomes more eccentric. This is physically reasonable. A highly eccentric orbit brings the neutron star much closer to its companion during pericentre passage, where the tidal field is strongest. The strong dependence of the tidal contribution on the orbital separation makes the pericentre distance particularly important. Consequently, the tidal precession is evaluated for different values of pericentre distances $r_p$.

For the present study, a sequence of neutron-star configurations is generated by solving the TOV equations using the polytropic EOS 
\begin{equation}
    P=K\rho^\Gamma
\end{equation}
with \[ K=200,\qquad \Gamma=2. \]
For each equilibrium configuration, we determine M, R, C and $k_2$. These quantities are then used to calculate $ \Delta\phi_{\rm tidal} $ for different orbital separations. The resulting sequence provides a direct connection between the internal structure of the neutron star and its orbital dynamics.

\begin{figure}[htbp]
\centering
\includegraphics[width=1.05\linewidth]{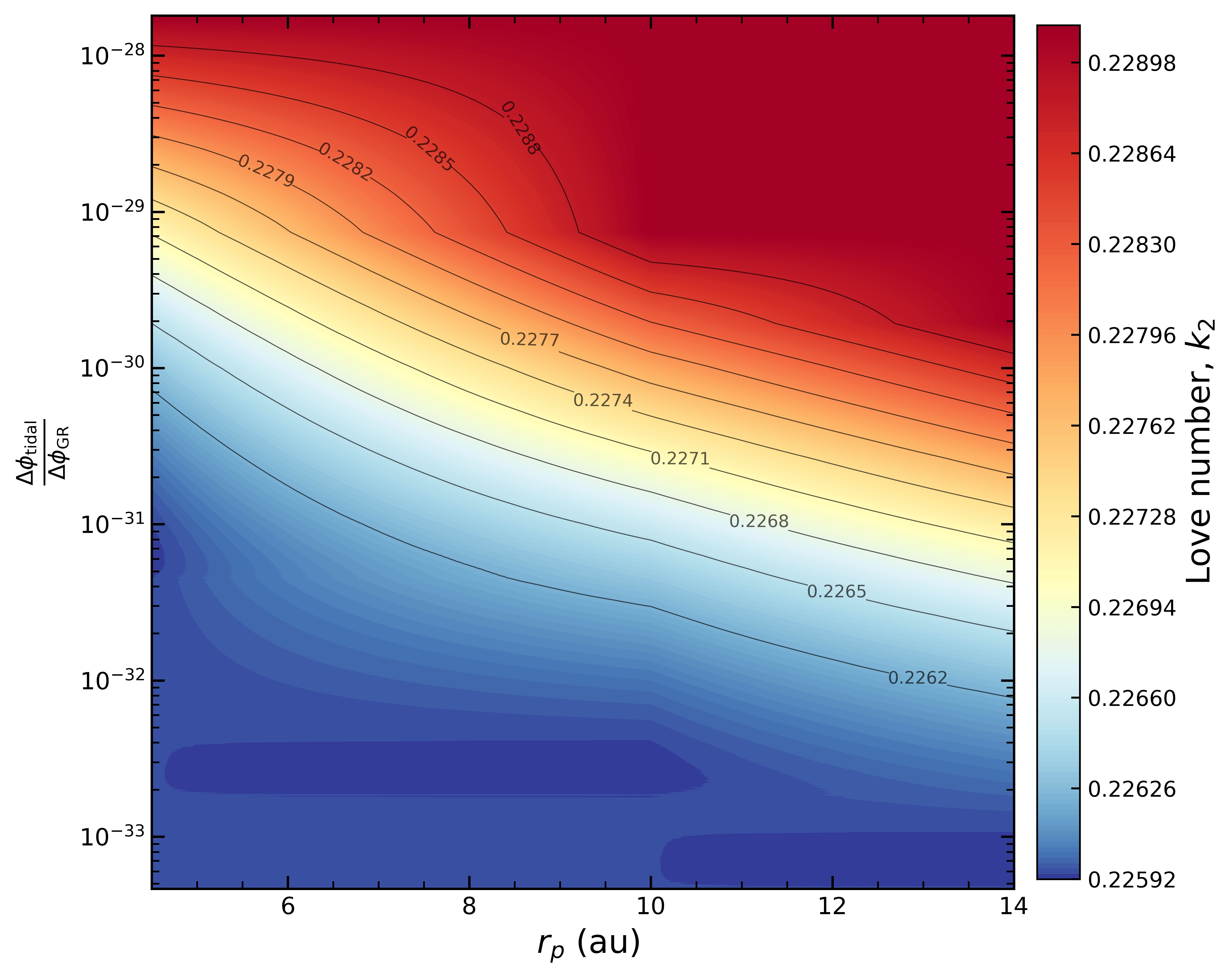}
\caption{Ratio of the tidal precession to the general-relativistic  precession per orbit, $\Delta\phi_{\rm tidal}/\Delta\phi_{\rm GR}$, as a function of the pericentre distance $r_p$ for neutron stars for different $k_2$.}
\label{fig:k2_mass}
\end{figure}

\section{Results and Discussion}
Figure~\ref{fig:tide_compact}  illustrates the dependence of tidal precession on stellar compactness for four pericentre distances: $r_p$ = 0.009au, 4.5au, 12au and 100au. The smallest distance represents an orbit close to the Schwarzschild radius, while \(r_p = 4.5\,\mathrm{au}\) corresponds to the adopted estimate of the innermost stable circular orbit of Sgr A* (see Equation 1 from \cite{r24}). The distances \(r_p = 12\) and \(100\,\mathrm{au}\) correspond to the pericentre distances adopted for S-301 and S-2, respectively. In all four cases, tidal precession decreases with increasing compactness, with a steeper decline towards the upper end of the plotted range. Smaller pericentre distances consistently yield larger tidal precession, and the ordering of the curves remains unchanged throughout. These results highlight the strong dependence of the calculated tidal angular shift on the distance of closest approach.

Figure~\ref{fig:rp_vs_C_a}  shows that the tidal angular shift is greatest when both the compactness \(C\) and the pericentre distance \(r_p\) are small. At fixed compactness, the shift decreases rapidly with increasing pericentre distance, reaching its largest values near \(r_p = 4.5\,\mathrm{au}\). By comparison, the reference lines at \(r_p = 10\) and \(12\,\mathrm{au}\) lie predominantly within the lowest colour band. At fixed pericentre distance, increasing compactness also reduces the shift. The largest shifts are concentrated in the lower-left region of the plot, particularly where \(C \le 0.02\) and \(r_p < 6\,\mathrm{au}\), with a maximum displayed value of approximately \(3.5 \times 10^{-18}\,\mu\mathrm{as}\). The enhancement at smaller pericentre distances is consistent with stronger tidal interactions during closer encounters, while the decrease with increasing compactness indicates a weaker tidal response for more compact objects within the adopted model. Even under the most favourable conditions shown, however, the predicted angular signal remains extremely small.

Figure~\ref{fig:MR} illustrates the relationship between tidal precession, compactness, and the tidal Love number \(k_2\). At fixed compactness, larger values of \(k_2\) correspond to greater tidal precession. Along contours of constant \(k_2\), precession generally increases with compactness, although this dependence weakens towards the higher-compactness end of the plotted range. Conversely, at fixed precession, \(k_2\) generally decreases as compactness increases. Increasing the pericentre distance from \(4.5\) to \(10\,\mathrm{au}\) substantially reduces the predicted tidal precession. Nevertheless, the values remain extremely small in both cases, spanning approximately $10^{-21}$ - $10^{-18}$ $\mu as$ in the first panel and \(10^{-23}\)–\(10^{-20}\,\mu\mathrm{as}\) in the second.

Figure~\ref{fig:k2_mass} shows that, at fixed \(k_2\), the ratio of tidal to general relativistic (GR) precession decreases with increasing pericentre distance, as indicated by the downward-sloping contours. The tidal contribution therefore becomes progressively less significant relative to GR precession for more distant encounters. At fixed pericentre distance, larger values of \(k_2\) generally correspond to larger precession ratios. Although the Love number varies only over the narrow interval \(k_2 \approx 0.2259\)–\(0.2291\), the plotted ratio spans approximately \(10^{-33}\)–\(10^{-28}\). Thus, within the adopted model and normalization, GR precession overwhelmingly dominates the tidal contribution throughout the displayed parameter range. These results indicate that the tidal contribution calculated using the Tolman–Oppenheimer–Volkoff (TOV) stellar model is exceedingly small, making its direct detection through neutron-star orbital precession exceptionally challenging. 

The extremely small tidal precession arises primarily from the fifth-power dependence on the ratio of stellar radius to pericentre distance, $(R/r_p)^5$. For neutron stars, this ratio is sufficiently small that its suppression outweighs the enhancement associated with the black-hole-to-stellar mass ratio. Consequently, direct detection of the neutron-star tidal contribution through orbital precession around Sgr A* appears exceptionally challenging within the adopted model. These findings establish the negligible role of tidal corrections in the calculated precession over the investigated parameter range.

\begin{acknowledgements}

The research of P.C. Lalremruati and 
N.
Malsawmtluangi is funded by the Anusandhan National Research Foundation (ANRF), Government of India, under 
Grant No. ANRF/IRG/2024/000836/PS.
\end{acknowledgements}

\end{document}